\documentclass[journal]{IEEEtran}
\usepackage{array}
\IEEEoverridecommandlockouts
\usepackage{cite}
\usepackage{amsmath,amssymb,amsfonts}
\usepackage{algorithm}
\usepackage{algpseudocode}
\usepackage{graphicx}
\usepackage{amsmath}
\usepackage{amsthm}
\usepackage{textcomp}
\usepackage{xcolor}
\usepackage{amsmath}
\usepackage{float} 
\usepackage{comment}
\usepackage{subfigure} 
\usepackage{color}
\usepackage{bm}
\usepackage{esvect}

\usepackage[top=0.73in, bottom=0.68in, left=0.58in, right=0.58in]{geometry}

\def\BibTeX{{\mathrm B\kern-.05em{\sc i\kern-.025em b}\kern-.08em
    T\kern-.1667em\lower.7ex\hbox{E}\kern-.125emX}}

\begin{document}
\title{\huge
Active-IRS-Aided Wireless Communication:\\ Fundamentals, Designs and Open Issues}
\author{Zhenyu Kang, Changsheng You, 
	 and Rui Zhang
	   \thanks{
	   Zhenyu Kang is with National University of Singapore;    
	   Changsheng You is with Southern University of Science and Technology;     
	   Rui Zhang is with School of Science and Engineering, Shenzhen Research Institute of Big Data, The Chinese University of Hong Kong, Shenzhen, and the National University of Singapore. (Corresponding author: Rui Zhang and Changsheng You).

         This work was supported in part by the 2022 Stable Research Program of Higher Education of China under Grant 20220817144726001, the National Natural Science Foundation of China under Grant 62201242, the Ministry of Education, Singapore under Award T2EP50120-0024, the Advanced Research and Technology Innovation Centre (ARTIC) of National University of Singapore under Research Grant R-261-518-005-720, and The Guangdong Provincial Key Laboratory of Big Data Computing.
	   }
	   }
\maketitle

\begin{abstract}
Intelligent reflecting surface (IRS) has emerged as a promising technology to realize smart radio environment for future wireless communication systems. Existing works in this line of research have mainly considered the conventional passive IRS that reflects wireless signals without power amplification, while in this article, we give an overview of a new type of IRS, called \emph{active} IRS, which enables simultaneous signal reflection and amplification, thus significantly extending the signal coverage of passive IRS. We first present the fundamentals of active IRS, including its hardware architecture, signal and channel models, as well as practical constraints, in comparison with those of passive IRS. Then, we discuss new considerations and open issues in designing active-IRS-aided wireless communications, such as the reflection optimization, channel estimation, and deployment for active IRS, as well as its integrated design with passive IRS. Finally, numerical results are provided to show the potential performance gains of active IRS as compared to passive IRS and traditional active relay.
\end{abstract}
\section{Introduction}

To accommodate assorted emerging new applications, e.g., holographic communication, extended reality, and tactile Internet, the next-generation wireless networks need to meet more stringent performance requirements than today's fifth-generation (5G) networks, such as unprecedentedly high throughput, super-high reliability, ultra-low latency, extremely low power consumption, etc. 
These requirements, however, may not be fully achieved by existing 5G technologies such as massive multi-input multi-output (MIMO) and millimeter wave (mmWave) communications cost-effectively \cite{9040264}, since they enhance communication performance with significantly increased energy consumption and hardware cost.
Moreover, the performance of 5G networks is still fundamentally constrained by the random and uncontrollable wireless channels due to various propagation impairments and user mobility.
To address the above issues, a promising new technology, called \emph{intelligent reflecting surface} (IRS), has been proposed to dynamically control the radio propagation environment and improve the wireless network performance in a cost-effective manner \cite{9326394}.
Specifically, IRS is a planar meta-surface consisting of massive low-cost passive reflecting elements, which are able to reflect the incident signals with individually tuned phase shifts and/or amplitudes.
As such, the reflected signals by all reflecting elements can be added constructively for enhancing the signal power in desired direction (so-called passive beamforming) or destructively for mitigating undesired interference.

\begin{figure*}[t]
\centerline{\includegraphics[width=5in]{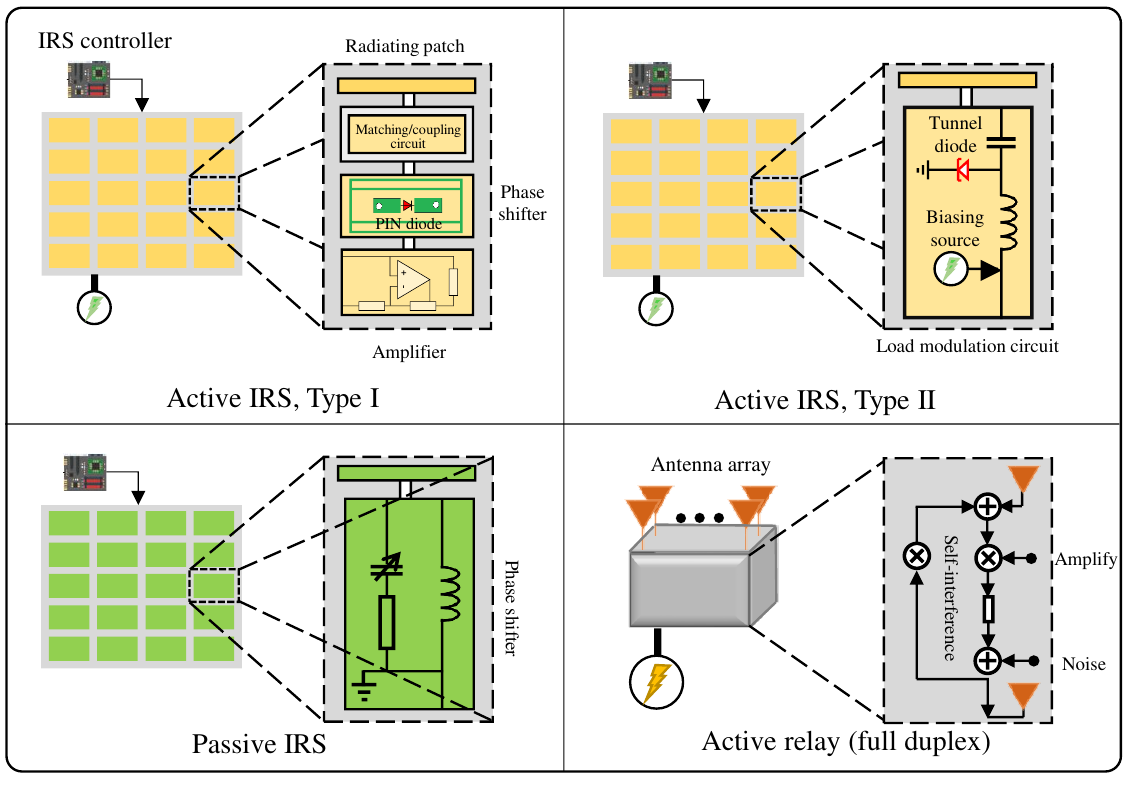}}
\caption{Hardware architectures of active IRS, passive IRS, and active relay.}\label{sysmod}
\end{figure*}

However, conventional IRS is mounted with fully passive reflecting elements (referred to as \emph{passive} IRS), thus facing several issues in practice.
In particular, the reflected signal by IRS suffers severe \emph{product-distance path loss}, i.e., the cascaded path loss in the transmitter-IRS-receiver link is the product of those in the transmitter-IRS and IRS-receiver links, respectively. 
This fundamentally constrains the signal power at the receiver and hence the effective coverage of each IRS.
Two practical methods to resolve this issue have been proposed, which are, respectively, deploying more passive elements at the IRS to enhance its passive beamforming gain and placing the passive IRS closer to either the transmitter or the receiver to reduce the cascaded channel path loss.
However, these methods may not be applicable in practice since 1) equipping IRS with massive reflecting elements or placing it too close to the transceivers will impose great difficulty in practical IRS deployment, and 2) more reflecting elements generally incur higher training overhead for their corresponding channel estimation and higher complexity for optimizing their reflection coefficients in real time\cite{9326394}.

To tackle this problem, a new type of IRS, called \emph{active} IRS, has been recently proposed to enable simultaneous signal reflection and \emph{amplification}, thus more effectively compensating the severe cascaded path loss as compared to passive IRS without signal amplification \cite{9377648,9734027,daiActive}.
As shown in Fig. \ref{sysmod}, active IRS comprises a number of meta-atoms with reflection-type amplifiers. Compared to conventional passive IRS, active IRS can simultaneously alter the signal's phase and amplify its amplitude to enhance the signal power at the receiver with moderately higher power consumption and hardware cost.
Moreover, the amplitude control with power amplification by active IRS provides more degrees-of-freedom (DoFs) to achieve finer-grained reflective beamforming than passive IRS.
In addition, the reflection-type amplifiers of active IRS can directly amplify incoming signals in a full-duplex manner.
Compared to traditional half-duplex or more advanced full-duplex active relays, reflection-type amplifiers of active IRS are more cost- and energy-efficient as they do not require expensive and power-hungry radio frequency (RF) chains.
The above appealing features of active IRS offer the great potential to significantly enhance the performance of future sixth-generation (6G) wireless applications, such as indoor localization, autonomous vehicle, by enhancing desired signal power, suppressing undesired interference, and improving transmission reliability, etc.
Despite the above advantages, active IRS also brings new design issues, which need to be studied to evaluate its practical performance gain over passive IRS and traditional active relay. 
For example, due to signal amplification, the reflecting components of active IRS generate non-negligible amplification noise, which results in degraded signal-to-noise ratio (SNR) at the receiver and thus calls for efficient techniques to balance the trade-off between signal and noise amplification by active IRS. 
Moreover, as active IRS usually has limited power supply in practice, it is of paramount importance to design its size (number of reflecting elements) and placement for optimizing the communication performance.

Motivated by the above, this article aims to provide a comprehensive overview of active-IRS-aided wireless communications. 
To this end, we first present the fundamentals of active IRS, including its hardware architecture, signal and channel models, as well as practical constraints. Next, we discuss new considerations and open issues in designing active-IRS-aided communication systems, such as the reflection optimization, channel estimation, and deployment for active IRS, as well as how to efficiently integrate it with conventional passive IRS to achieve superior performance than stand-alone active/passive IRS only.
Finally, numerical results are presented to show the potential performance gains of active IRS as compared to conventional passive IRS and active relay.

\allowdisplaybreaks[4]
\section{Fundamentals of Active IRS}

In this section, we present the fundamentals of active IRS, and compare it with conventional passive IRS and active relay for wireless relaying.

\subsection{Hardware Architecture}%ref 4 5 6 24

\begin{table*}[ht]\tiny
\caption{Comparisons of active IRS, passive IRS, and active relay.}\label{tab_I}
\centering
\resizebox{14cm}{!}{%
\begin{tabular}{|c|c|c|c|}
\hline
 & \textbf{Active IRS} & \textbf{Passive IRS} & \textbf{Active relay} \\ \hline
\textbf{Hardware cost} & Low & Very low & High \\ \hline
\textbf{Power consumption} & Moderate & Very low & High \\ \hline
\textbf{Operation mechanism} & Reflect and amplify & Reflect & Receive and amplify \\ \hline
\textbf{Duplex mode} & Full duplex & Full duplex & Full/half duplex \\ \hline
\textbf{Amplification noise} & Yes & No & Yes \\ \hline
\end{tabular}%
}
\end{table*}

As shown in Fig. \ref{sysmod}, active IRS mainly consists of a smart controller and a number of properly designed reflecting elements/meta-atoms. 
Specifically, the IRS controller provides appropriate stimulus to adjust the reflection coefficients, i.e., amplitude and phase shift, of each reflecting element via e.g., a field programmable gate array (FPGA)-based control board. 
In addition, it is also responsible for communicating with the base station (BS)/access point (AP) through dedicated wireless control links.
On the other hand, each active-IRS reflecting element can reflect the incident signal with desired phase shift and amplitude adjustment based on the electromagnetic (EM) scattering principle. 
This can be achieved, for example, by the cascaded amplifying and phase shifting (APS) circuits \cite{6019001,lonvcar2019ultrathin}, or load modulation circuits \cite{8403249,7920385}.
In Fig. \ref{sysmod}, we illustrate two typical hardware architectures for implementing active-IRS reflecting elements.
The first one (type I) is implemented by the APS circuit, which is the integration of an operational amplifier and a phase shifter \cite{6019001,lonvcar2019ultrathin}.
Specifically, triggered by a biasing power source, the operational amplifier amplifies the weak electric signal with desired amplification factor.
Meanwhile, the phase shifting circuit, e.g., positive-intrinsic-negative (PIN) diode, varactor diode \cite{6019001}, connected to the amplifier enables flexible phase modulation to the reflected signal.
In contrast, the second implementation of active-IRS reflecting element (type II) can be achieved by load modulation circuits based on e.g., tunnel diodes \cite{8403249} and complementary metal-oxide semiconductor (CMOS) \cite{7920385}. 
Taking the tunnel diode \cite{8403249} as an example, by imposing proper bias voltage, it can operate in the region with negative differential resistance such that the reflecting element has an active load impedance (negative resistance).
As such, the reflection amplitude of incident signals can be increased by converting the direct current (DC) power into the RF power, while the phase shift is adjusted by tuning the circuit capacitance.

\subsection{Signal and Channel Models}

The signal and channel models for active IRS are more complicated than those for passive IRS.
First, different from the noise-free passive IRS, the noise introduced by active IRS is composed of both dynamic noise and static noise. 
Specifically, the dynamic noise is related to the input noise which is determined by the source resistance of the antenna and inherent device noise; while the static noise is independent of the transmit/amplification power, which is usually negligible compared to the dynamic noise.
Second, the signal at the receiver via active IRS is superposed by the desired signals over both the cascaded transmitter-IRS-receiver and direct transmitter-receiver links, the amplification noise generated at each reflecting element over the IRS-receiver link, and the additive noise at the receiver. 
Let $M$ denote the number of active-IRS elements and $x$ denote the transmitted signal.
Then the received signal $y$ at the receiver can be mathematically expressed as $y = \left(\mathbf{h}^H\mathbf{\Psi}\mathbf{g}+t\right)x+\mathbf{h}^H\mathbf{\Psi}\mathbf{z}_\mathrm{I}+z_0$, where $\mathbf{h}^H$, $\mathbf{g}$, and $t$ denote the equivalent baseband channels of the IRS$\to$receiver, transmitter$\to$IRS, and transmitter$\to$receiver links; $\Psi \triangleq \operatorname{diag}\left(\alpha_1, \cdots, \alpha_M\right) \operatorname{diag}\left(e^{\jmath \phi_1}, \cdots, e^{\jmath \phi_M}\right)$ is the active-IRS reflection matrix with $\alpha_m$ and $\phi_m$ representing the amplification factor and phase shift of the $m$-th reflecting element, respectively;
$\mathbf{z}_\mathrm{I}$ and $z_0$ denote the amplification noise induced by active IRS and the additive noise at the receiver. 

\subsection{Practical Constraints}\label{Sec2c}

Active IRS usually has smaller amplification power than active relay, which thus limits its maximum amplification factors at the reflecting elements in practice. 
Generally speaking, there are two practical amplification power models applicable to active IRS.
In the first model, the amplification factors of all reflecting elements are subject to a total power constraint as they share a common power source \cite{9377648}. 
In this case, the total amplification power can be allocated to reflecting elements based on the corresponding IRS-related channels to achieve optimal performance.
While for the second model, in contrast, each active-IRS reflecting element is attached with a separate power source with independent power control.
As such, the per-element maximum amplification factor is constrained by the \emph{individual} per-element amplification power. 

Moreover, in practice, other hardware constraints should be considered for modeling active IRS. 
First, the reflection amplitude based on the negative resistance circuit is generally phase-dependent, because both the amplification factor and phase shift are functions of the effective capacitance and resistance of each reflecting element.
Second, the load impedance of the active-IRS element based on tunnel diode is determined by the effective capacitance and resistance.
As a result, the maximum amplification factor of the element is constrained by its inherent load impedance. This load impedance cannot be tuned to be arbitrarily large, regardless of the power supply.
Last but not least, the resolutions of active-IRS phase shift and amplification factor depend on the specific hardware implementation, which need to take discrete values in practice.

\subsection{Comparison with Passive IRS and Active Relay}% ref 35

In Table \ref{tab_I}, we compare the main hardware characteristics and implementations of active IRS with its two counterpart technologies for wireless relaying, namely, passive IRS and active relay, while their main differences in communication designs will be discussed in more details in  Section \ref{design}.

First, in terms of hardware cost and power consumption, active IRS surpasses passive IRS due to the additional reflection-type amplifiers, while it falls below active relay since the latter requires power-hungry and costly RF chains for signal processing and amplification. 

Next, compared to passive IRS, which can only tune the phase of the reflected signal, active IRS can tune both the phase and amplitude of the reflected signal, thus providing a higher DoF for more flexible beamforming, e.g., narrower beam and larger coverage (see Fig. \ref{network}(a)).
However, its beamforming performance is generally inferior to half-/full-duplex active relay, since active IRS only reflects and amplifies signals in the RF band with an equivalent diagonal baseband reflection matrix (i.e., $\Psi$), while active relay with multiple antennas have more sophisticated signal processing capability with typically a full-rank MIMO signal processing matrix at the baseband.

Third, wireless communication systems aided by active IRS generally attain higher spectral efficiency than those aided by passive IRS and half-duplex active relay. 
On one hand, active IRS enables power amplification when reflecting signals, thus leading to a higher received SNR at the receiver than passive IRS. 
On the other hand, practical active relay usually operates in the half-duplex mode\footnote{Full-duplex relay requires higher hardware and implementation cost than half-duplex relay due to the need to eliminate the loop-back self-interference.}, thus suffering considerable rate performance loss compared to full-duplex active IRS.

Last, the self-interference of full-duplex relays and that of active IRSs are also different. Specifically, the self-interference of full-duplex relays is much stronger than that of active IRSs since they use different types of power amplifiers (RF-chain-based versus reflection-type). In addition, due to signal processing delay, the self-interference in full-duplex relays involve transmitted symbols that are different from the incident symbols, thus resembling the inter-symbol interference. In contrast, in the ideal case, active IRSs have negligible self-interference, since its processing delay is typically in nanosecond and hence the incident and reflected signals carry the same symbol \cite{daiActive}. Nevertheless, in practice, due to the non-ideal inter-element isolation, some of the reflected signals may be received again by the active elements, resulting in the loop-back self-interference. This should be well suppressed in the hardware design and may need to be considered in the active-IRS reflection optimization.

\section{New Design Considerations and Challenges}\label{design}

In this section, we present new considerations and open issues in designing active-IRS-aided wireless communications. Moreover, promising solutions for tackling these issues are also proposed to inspire future work.

\subsection{Active-IRS Reflection Optimization}

One key challenge in active-IRS-aided communication design is how to optimize the reflection coefficients of active IRS for achieving the optimal beamforming and SNR performance. 
To address it, several practical considerations need to be taken into account such as the phase-amplitude dependent control, maximum amplification factor, discrete phase shifts/amplitude levels (cf. Section \ref{Sec2c}).
These considerations usually result in non-convex constraints over the reflection coefficients and hence non-convex optimization problems \cite{daiActive}. 
In addition, as the amplification factor of active-IRS element can be controlled, the resulting design variables are significantly more than those of passive IRS for which the reflection amplitude is usually set as one or its maximum value.
Several algorithms have been proposed in the literature to solve the above problem, see, e.g., \cite{9377648,9734027,daiActive,6019001}.
For example, the alternating optimization (AO) method can be applied to iteratively optimize the amplification factor and phase shift, until the convergence is achieved. 
However, the performance of AO critically depends on its initialization, which, if not properly designed, can result in considerable performance loss. 
Existing works, e.g., \cite{9377648,9734027,daiActive,6019001}, have usually assumed ideal reflection coefficients with continuous and independent phase-shift and amplitude control. However, these ideal assumptions cannot be realized in practice due to hardware limitations. 
Therefore, future research is needed to develop efficient algorithms that can handle more practical constraints.

Several results on the active-IRS reflection design under different channel conditions and/or system setups have been reported in the literature, e.g., \cite{9734027,9530750}.
Taking the single-user case with active or passive IRS as an example, it was shown that the reflected signals over different IRS elements should be phase-aligned such that they can be constructively added at the receiver \cite{9530750}.
Moreover, the amplification factor should be set to the maximum available value for compensating the cascaded channel path loss.
Next, for the general multi-user case, there exists a new trade-off between signal power enhancement and interference mitigation in designing the active-IRS reflection coefficients given a total power budget constraint.
In this case, the amplification factors of reflecting elements should be properly set, which may not necessarily equal to their maximum values.
Generally speaking, active IRS is expected to achieve better interference mitigation performance than passive IRS, since its amplitude control endows more DoFs to eliminate the undesired interference more effectively.

An important performance metric in the active-IRS reflection design is how the received SNR scales with the number of reflecting elements, $M$, when $M$ is asymptotically large.
First, consider the case with a total power constraint over all active-IRS elements. 
It was reported in \cite{9377648} that the received SNR with an active IRS linearly increases with $M$ when $M$ is sufficiently large.
This is expected since the signal power at the receiver quadratically increases with $M$ thanks to the IRS beamforming gain, while the amplification noise power also increases linearly with $M$, thus leading to the SNR scaling order of $M$.
However, it is worth noting that although active IRS has a smaller SNR scaling order than passive IRS without amplification noise, i.e., $\mathcal{O}\left(M\right)$ versus $\mathcal{O}\left(M^2\right)$ \cite{9326394}, it provides superior rate performance than passive IRS when the number of reflecting elements is moderate and/or its amplification power is sufficiently large. This is because active IRS provides an appealing power amplification gain for significantly enhancing the received SNR, while passive IRS has a limited passive beamforming gain when the number of reflecting elements is not large.
Next, consider the case with fixed per-element power for each active-IRS element.
In this case, the total power budget linearly increases with the number of reflecting elements. This opens up a new research direction to investigate the performance of active-IRS with a practical fixed per-element power in various scenarios.
As such, the active IRS is expected to achieve additional amplification power gain in the order of $M$, hence yielding the same received SNR scaling order with the conventional passive IRS, i.e., $\mathcal{O}\left(M^2\right)$ (such SNR comparisons will be numerically shown in Fig. \ref{case2}).

\begin{figure*}[t]
\centerline{\includegraphics[width=6.5in]{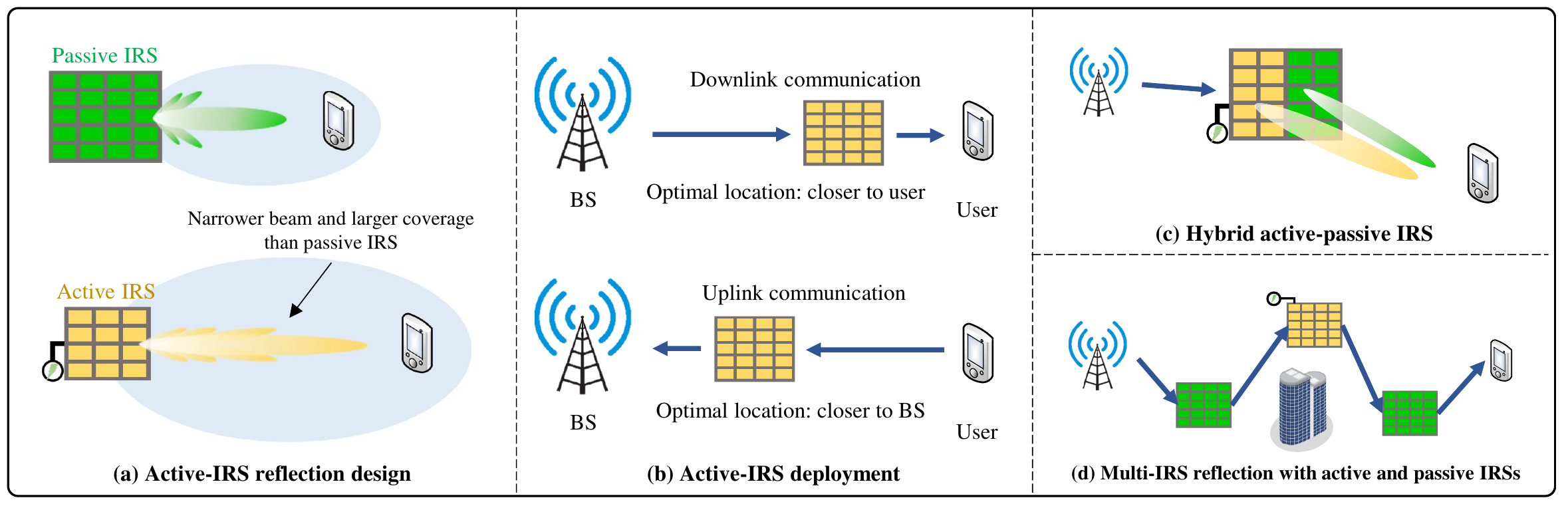}}
\caption{New design considerations for active-IRS-aided wireless communications.}\label{network}
\end{figure*}

\subsection{Active-IRS Channel Estimation}

Active-IRS reflection design requires accurate channel state information (CSI) of IRS-related channels for fully exploiting its beamforming and amplification gains.  
In particular, the channel estimation for active-IRS-aided communications may incur a large channel training overhead when the number of reflecting elements is large since more channel coefficients are to be estimated. This issue becomes particularly severe for short packet communications.
Compared with passive IRS, the CSI acquisition for active IRS is generally more challenging due to the following reasons.
First, the active-IRS reflection design generally requires explicit CSI of the separate transmitter-IRS and IRS-receiver links due to the new consideration of amplification noise \cite{9530750}, which greatly differs from the passive-IRS case that requires cascaded CSI only.
Specifically, although the phase shift optimization can be based on cascaded CSI, controlling the amplification factor of active IRS is generally dependent on the incident signal power and power budget, which in turn depends on the transmitter-IRS CSI.
Second, in active-IRS channel estimation, setting the amplification factor to meet the power constraint is difficult.
This is because it requires the knowledge of the incident signal power, which also depends on the transmitter-IRS CSI; however, it has yet to be estimated.
Last, the amplification noise also affects the channel estimation performance, while this issue does not exist in the channel estimation for passive IRS.

In the following, we introduce two practical active-IRS channel estimation approaches, with and without IRS-mounted sensors, respectively.
The first approach requires low-cost sensors installed on the active IRS, which are interleaved with the reflecting elements.
Equipped with these sensors that can receive signals sent by the BS/users, active IRS can estimate the BS-IRS and IRS-user channels separately by exploiting the strong channel correlations between the IRS elements and sensors. 
It is worth noting that active IRS in general requires much fewer reflecting elements than passive IRS for achieving comparable rate performance, thus it is easier to place reflecting elements with sensors close to each other to achieve high channel correlation.
Second, for active IRS without sensors installed, one promising approach is to combine the separate and cascaded channel estimation methods.
Specifically, the BS first estimates the BS-IRS CSI based on the reflected signal over the BS-IRS-BS link, i.e., by wireless sensing of the BS.
Next, the BS estimates the cascaded BS-IRS-user CSI based on the pilot signals from the user, and then resolves the IRS-user CSI based on the estimated cascaded CSI and BS-IRS CSI.
To facilitate the estimation of IRS-user CSI, there are two practical methods for the amplification factor control. The first method is setting an equal IRS reflection amplitude over all reflecting elements based on the received signal powers (with sensors). The second method is simply setting the amplification factor equal to one (without sensors), similar to the case of passive IRS.

\subsection{Active-IRS Deployment}\label{deploy}

For the active-IRS deployment design, several factors need to be considered, such as the locations of users, uplink or downlink communication, number of reflecting elements, amplification power budget, amplification noise power, etc. 

First, consider the active-IRS location optimization for the single-user case.
In this case, it has been shown in \cite{9530750} that active IRS should be properly deployed between the transmitter and receiver to balance the large-scale path loss of the transmitter-IRS and IRS-receiver channels. 
In particular, with a small amplification power or a large number of reflecting elements, active IRS should be deployed closer to the receiver; otherwise, the amplification factors are too small to compensate the amplification noise.
This is different from the passive-IRS case where the IRS should be deployed near either the transmitter or receiver to minimize the cascaded path loss.
Besides, the optimal active-IRS deployment locations are different in the uplink and downlink for two-way communications in general, and thus their performance should be balanced by choosing an appropriate location for active IRS (see Fig. \ref{network}(b)). 

Second, for the multi-user case, the active-IRS deployment design is more complicated.
On one hand, it needs to strike the performance balance among different users.
On the other hand, the active IRS may not be deployed at the BS/AP side as in the passive-IRS case, since it also needs to balance the powers of desired signal and amplification noise.
To tackle this problem, one efficient and low-complexity approach is by leveraging machine learning (ML) and geometry-based optimization methods to deploy the active IRS at a location with good statistical channel conditions. 
Besides active-IRS location, the reflection gain of active IRS is also affected by its rotation angle.
As shown in \cite{9722711}, considerable rate performance improvement can be attained by properly rotating the IRS without changing its locations. Thus, the IRS rotation should be carefully tuned to maximize the number of served users and IRS-user channel gains.
Moreover, compared with single-IRS deployment, the multi-active-IRS deployment is generally more challenging.
In particular, there exists a fundamental trade-off between the deployment cost and communication performance.
As such, besides deployment locations, the number of active IRSs needs to be carefully chosen to minimize the total deployment cost, while achieving the communication performance requirements.

\subsection{Integrated Design with Passive IRS}

While the existing works mainly considered the wireless communication systems aided by active or passive IRS only, it has been recently shown that the integrated design with both active and passive IRSs has the potential to achieve their combined advantages at the same time \cite{hybrid,hybridIRS,min}, as discussed below.

First, it was shown in \cite{hybrid} that given the total deployment budget and amplification power, a hybrid IRS with both passive and active elements (see Fig. \ref{network}(c)) can achieve better communication performance than the conventional IRS with active or passive elements only.
Specifically, there generally exists a trade-off between deploying more active elements to achieve the appealing power amplification gain and deploying more passive elements to attain a higher beamforming gain. This thus leads to the new design issue of active versus passive elements allocation \cite{hybrid}.
Generally speaking, when the number of reflecting elements is small and/or the amplification power is large, more active elements should be deployed to harness the power amplification gain; otherwise, deploying more passive elements is expected to achieve better performance thanks to their higher-order beamforming gain \cite{hybrid}.

Another type of the integrated design is deploying both \textit{stand-alone} active and passive IRSs in the wireless network to improve the communication
performance, which are equipped with either active or passive elements only, as illustrated in Fig. \ref{network}(d).
The advantages are twofold. First, deploying multiple cooperative IRSs provides the opportunity to construct a multi-hop reflection path between the transmitter and receiver to bypass environmental obstacles \cite{9528043}. 
Second, compared with the conventional multi-passive-IRS network, adding a number of active IRSs in the network can substantially enhance the communication performance. 
This is because active IRSs can opportunistically amplify the reflected signal along the multi-reflection path, thus effectively compensating the severe end-to-end product-distance path loss.
Despite the advantages, the design of multi-active and multi-passive (MAMP) IRS aided communications is also more challenging \cite{hybridIRS}. 
For example, the multi-IRS beam routing path with both active and passive IRSs should be carefully devised to maximize the power amplification gain, and at the same time, minimize the accumulated amplification noise power arising from active IRSs.
Moreover, the numbers of active and passive IRSs should also be carefully chosen given limited deployment budget and their locations should be optimally designed \cite{min}, which deserve further studies.

\section{Numerical Results}

\begin{figure}[t]
\centerline{\includegraphics[width=3.3in]{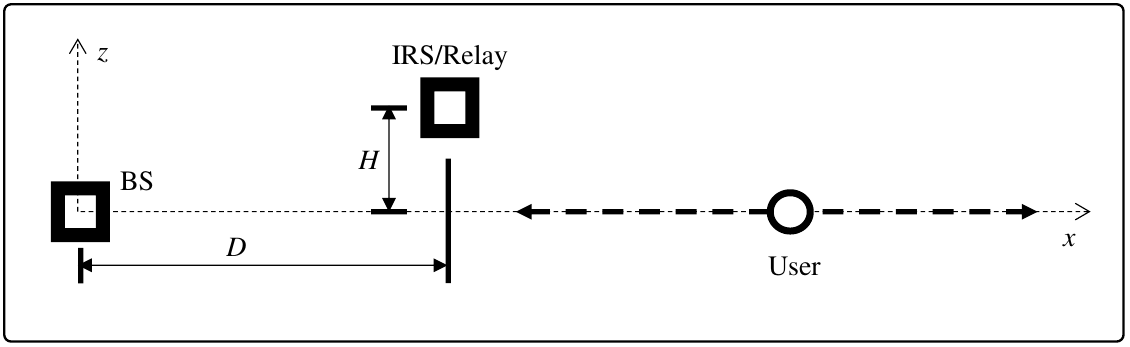}}
\caption{Simulation setup.}\label{setup}
\end{figure}

Numerical results are presented in this section to demonstrate the potential performance gains of active IRS as compared to passive IRS and active relay.
Specifically, we consider a typical point-to-point communication system as shown in Fig. \ref{setup}, where an IRS or a relay is deployed at an altitude of 2 meters (m) to assist the communication between a BS and a user (both equipped with single antenna), while the direct BS-user channel is assumed being blocked.
For each reflecting element, we consider the continuous and independently controlled phase shift and reflection amplitude (the amplitude is set to be one for passive IRS).
Moreover, the line-of-sight (LoS) channel model is assumed for all involved links, where the path loss exponent is 2 and the reference channel power gain at a distance of 1 m is $-30$ dB. Both powers of the receiver noise and per-element active-IRS amplification noise are set as $-80$ dBm.

\begin{figure}[t]
\centerline{\includegraphics[width=2.7in]{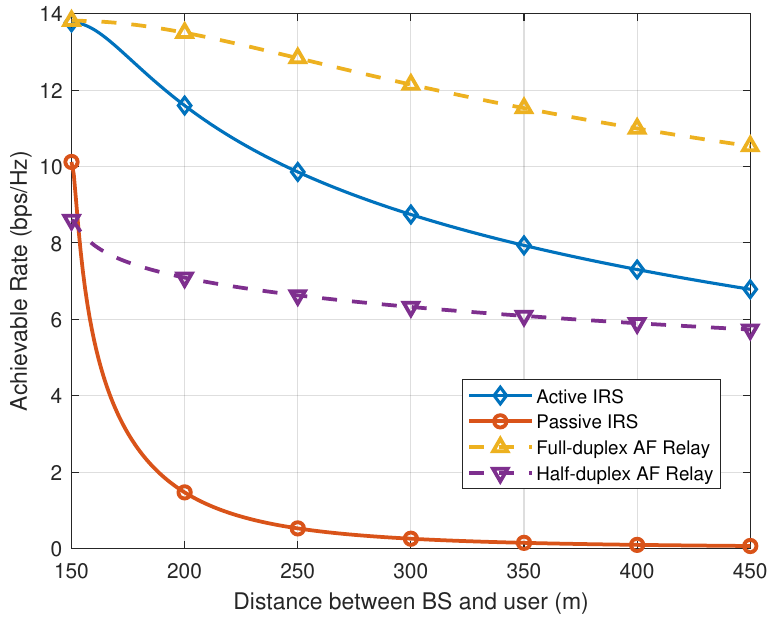}}
\caption{Achievable rate versus user location.}\label{case1}
\end{figure}

Fig. \ref{case1} shows the user's maximum achievable rates versus the BS-user distance in four wireless communication systems aided by:
1) an active IRS with 128 reflecting elements and a total amplification power of 10 dBm;
2) a passive IRS with 128 reflecting elements;
3) an ideal full-duplex amplify-and-forward (AF) relay with perfect self-interference cancellation and a total amplification power of 15 dBm;
and 4) a half-duplex AF relay with optimal time division and a total amplification power of 15 dBm.
To ensure fair comparison, we assume the same total power budget for the above four systems including the transmit power of the BS and the amplification power of the active IRS/relay.
It is observed that given the same BS-user distance, active IRS always achieves a higher rate than passive IRS and half-duplex relay.
This is expected since active IRS provides an additional power amplification gain compared to passive IRS, and it achieves the full-duplex gain compared to half-duplex relay.
Besides, active IRS is inferior to full-duplex active relay due to its limited signal processing capability and less power supply than active relay.

\begin{figure}[t]
\centerline{\includegraphics[width=2.7in]{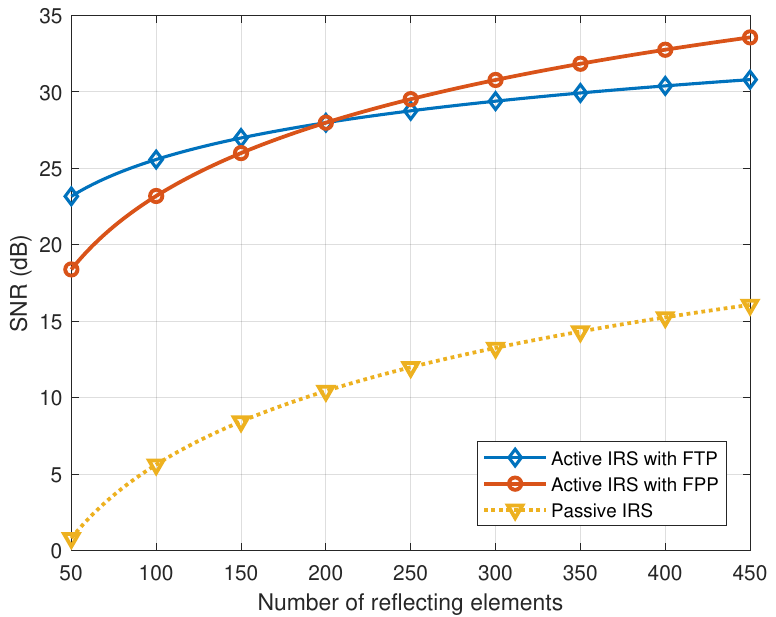}}
\caption{Received SNR versus number of reflecting elements.}\label{case2}
\end{figure}

Next, we show in Fig. \ref{case2} how the received SNR scales with the number of reflecting elements in three wireless communication systems aided by:
1) an active IRS with a fixed total power (FTP) of 0.2 W;
2) an active IRS with a fixed per-element power (FPP) of 1 mW;
and 3) a passive IRS.
For the above three systems, the corresponding IRS reflections are optimized for maximizing the received SNR.
First, it is observed that both active-IRS-aided systems achieve much higher received SNRs than the passive-IRS counterpart.
This is expected because active IRS provides both the beamforming gain and power amplification gain, while passive IRS only offers the beamforming gain.
Second, one can observe that with an increasing number of reflecting elements, the received SNR of active IRS with FPP is firstly lower than and then exceeds that with FTP. This is because in the latter case, the average per-element power decreases with the number of reflecting elements, thus yielding a lower received SNR scaling order.

\section{Conclusions}

In this article, we provide a comprehensive overview of the newly proposed active IRS that enables simultaneous signal reflection and amplification.
Specifically, we first present the fundamentals of active IRS, including its hardware architecture, signal and channel models, as well as practical constraints. 
Then, new considerations and open issues in designing active-IRS-aided wireless communications are discussed to shed new light to promising solutions, including the reflection optimization, channel estimation, and deployment for active IRS, as well as its integrated design with passive IRS.
Last, we numerically show the potential performance gains of active IRS over passive IRS and traditional half-duplex active relay, and the SNR scaling orders of active IRS under different power constraints versus passive IRS.

% Generated by IEEEtran.bst, version: 1.14 (2015/08/26)

\section*{Biographies}
\noindent {\bf Zhenyu Kang} (zhenyu\_kang@u.nus.edu) is a Ph.D. candidate with National University of Singapore, Singapore. 

\noindent {\bf Changsheng You} (youcs@sustech.edu.cn) received his Ph.D. degree from The University of Hong Kong. He is currently an Assistant Professor with Southern University of Science and Technology, China.

\noindent {\bf Rui Zhang} [F'17] (rzhang@cuhk.edu.cn) received the Ph.D. degree from Stanford University. He is now a Principal’s Diligence Chair Professor with The Chinese University of Hong Kong, Shenzhen, and also a Professor with National University of Singapore. He is a Fellow of the IEEE and the Academy of Engineering, Singapore.

\end{document}